\documentclass[runningheads,a4paper]{llncs}

\usepackage{amssymb}
\usepackage{graphicx}
\usepackage[T1]{fontenc}
\usepackage{multirow}
\usepackage{amsmath}
\usepackage{hyperref}

\usepackage{url}
\urldef{\mailsa}\path|{placzek.bartlomiej, marcin.bernas}@gmail.com|    
\newcommand{\keywords}[1]{\par\addvspace\baselineskip
\noindent\keywordname\enspace\ignorespaces#1}

\begin{document}

\mainmatter  %

\title{Detection of malicious data in vehicular ad-hoc networks for traffic signal control applications}

\titlerunning{Detection of malicious data in vehicular ad-hoc networks...}
\author{Bart\l{}omiej P\l{}aczek \and Marcin Bernas}
\authorrunning{B. P\l{}aczek, M. Bernas}
\institute{University of Silesia, Institute of Computer Science,\\
B\k{e}dzi\'nska 39, 41-200 Sosnowiec, Poland\\
\mailsa\\}

\toctitle{Detection of malicious data in vehicular ad-hoc networks for traffic signal control applications}
\tocauthor{Bart\l{}omiej P\l{}aczek, Marcin Bernas}
\maketitle

\begin{abstract}
Effective applications of vehicular ad hoc networks in traffic signal control require new methods for detection of malicious data. Injection of malicious data can result in significantly decreased performance of such applications, increased vehicle delays, fuel consumption, congestion, or even safety threats. This paper introduces a method, which combines a model of expected driver behaviour with position verification in order to detect the malicious data injected by vehicle nodes that perform Sybil attacks. Effectiveness of this approach was demonstrated in simulation experiments for a decentralized self-organizing system that controls the traffic signals at multiple intersections in an urban road network. Experimental results show that the proposed method is useful for mitigating the negative impact of malicious data on the performance of traffic signal control. \footnote{Preprint of: P\l{}aczek, B., Bernas, M.: Detection of malicious data in vehicular ad-hoc networks for traffic signal control applications. in Gaj, P., Kwiecien, A., Stera, P. (eds.) Computer Networks. CCIS, vol. 608, pp. 72-82, 2016. The final publication is available at \href{http://link.springer.com/chapter/10.1007/978-3-319-39207-3_7}{link.springer.com}}

\end{abstract}

\section{Introduction}

Vehicular ad-hoc networks (VANETs) facilitate wireless data transfer between vehicles and infrastructure. The vehicles in VANET can provide detailed and useful data including their positions, velocities, and accelerations. This technology opens new perspectives in traffic signal control and creates an opportunity to overcome main limitations of the existing roadside sensors, i.e., low coverage, local measurements, high installation and maintenance costs. The availability of the detailed data from vehicles results in a higher performance of the traffic signal control \cite{bbp_-1,bbp_0,bbp_1}. 

The VANET-based traffic signal systems have gained considerable interest in recent years. In this field the various solutions have been proposed that extend existing adaptive signal systems for isolated intersections \cite{bbp_2,bbp_3}. For such systems the data collected in VANET are used to estimate queue lengths and vehicle delays. On this basis optimal cycle length and split of signal phases are calculated. Similar adaptive approach was also used to control traffic signals at multiple intersections in a road network \cite{bbp_4,bbp_5}. Particularly advantageous for VANET-based systems is the self-organizing signal control scheme, which enables a decentralized optimization, global coordination of the traffic streams in a road network, and improved performance \cite{bbp_6}.

Effective VANET applications in traffic signal control require new methods for real-time detection of attacks that are based on malicious data. Injection of malicious data can result in significantly decreased performance o the traffic signal control, increased vehicle delays, fuel consumption, congestion, or even safety threats.

This paper introduces a method for the above mentioned applications, which can be used to detect malicious data. The considered malicious data are injected by vehicle nodes that perform Sybil attacks, i.e., create a large number of false vehicle nodes in order to influence the operation of traffic signals. The proposed method detects the malicious data by combining a model of expected driver behaviour with a position verification approach.

The paper is organized as follows. Related works are discussed in Section 2. Section 3 introduces the proposed method. Results of simulation experiments are presented in Section 4. Finally, conclusions are given in Section 5.

\section{Related works }

The problem of malicious nodes detection in VANETs has received particular attention and various methods have been proposed so far \cite{bbp_7}. The existing solutions can be categorized into three main classes: encryption and authentication methods, methods based on position verification, and methods based on VANET modelling.

In the encryption and authentication methods, malicious nodes detection is implemented by using authentication mechanisms. One of the approaches is to authenticate vehicles via public key cryptography \cite{bbp_8}. The methods that use public key infrastructure were discussed in \cite{bbp_9}. Main disadvantages of such methods are difficulties in accessing to the network infrastructure and long computational time of encryption and digital signature processing. Public key encryption and message authentication systems consume time and memory. Thus, bandwidth and resource consumption is increased in the public key systems. 

In \cite{bbp_10} an authentication scheme was proposed, which assumes that vehicles collect certified time stamps from roadside units as they are travelling. The malicious nodes detection is based on verification of the collected series of time stamps. Another similar method \cite{bbp_11} assumes that vehicles receive temporary certificates from roadside units and malicious nodes detection is performed by checking spatial and temporal correlation between vehicles and roadside units. These methods require a dense deployment of the roadside units. 

Position verification methods are based on the fact that position reported by a vehicle can be verified by other vehicles or by roadside units \cite{bbp_12}. The key requirement in this category of the methods is accurate position information. A popular approach is to detect inconsistencies between the strength of received signal and the claimed vehicle position by using a propagation model. According to the method introduced in \cite{bbp_13} signal strength measurements are collected when nodes send beacon messages. The collected measurements are used to estimate position of the nodes according to a given propagation model. A node is considered to be malicious if its claimed position is too far from the estimated one. In \cite{bbp_14} methods were proposed for determining a transmitting node location by using signal properties and trusted peers collaboration for identification and authentication purposes. That method utilizes signal strength and direction measurements thus it requires application of directional antennas. 

Xiao et al. \cite{bbp_15} and Yu et al. \cite{bbp_16} proposed a distributed method for detection and localization of malicious nodes in VANET by using verifier nodes that confirm claimed position of each vehicle. In this approach, statistical analysis of received signal strength is performed by neighbouring vehicles over a period of time in order to calculate the position of a claimer vehicle. Each vehicle has the role of claimer, witness, or verifier on different occasions and for different purposes. The claimer vehicle periodically broadcasts its location and identity information, and then, verifier vehicle confirms the claimer position by using a set of witness vehicles. Traffic pattern analysis and support of roadside units is used for selection of the witness vehicles. Yan et al. \cite{bbp_17} proposed an approach that uses on-board radar to detect neighbouring vehicles and verify their positions.

The modelling-based methods utilize models that describe expected behaviour of vehicle nodes in VANET. These methods detect malicious nodes by comparing the model with information collected from the vehicles. Golle et al. \cite{bbp_18} proposed a general model-based approach to evaluating the validity of data collected form vehicle nodes. According to this approach, different explanations for the received data are searched by taking into account the possible presence of malicious nodes. Explanations that are consistent with a model of the VANET get scores. The node accepts data that are consistent with the most scored explanation. On this basis, the nodes can detect malicious data and identify the vehicles that are the sources of such data. Another method in this category relies on comparing the behaviour of a vehicle with a model of average driving behaviour, which is built on the fly by using data collected from other vehicles \cite{bbp_19}. 

In \cite{bbp_20} a malicious data detection scheme was proposed for post crash notification applications that broadcast warnings to approaching traffic. A vehicle node observes driver's behaviour for some time after the warning is received and compares it with some expected behaviour. The vehicle movement in the absence of any traffic accident is assumed to follow some free-flow mobility model, and its movement in case of accident is assumed to follow some crash-modulated mobility model. On this basis the node can decide if the received warning is true or false.

A framework based on subjective logic was introduced in \cite{bbp_21} for malicious data detection in vehicle-to-infrastructure communication. According to that approach, all data collected by a vehicle node can be mapped to a world-model and can then be annotated with opinions by different misbehaviour detection mechanisms. The opinions are used not only to express belief or disbelief in a stated fact or data source, but also to model uncertainty. Authors have shown that application of the subjective logic operators, such as consensus or transitivity, allows different misbehaviour detection mechanisms to be effectively combined.

According to the authors' knowledge, the problem of malicious data detection for traffic signal control applications has not been studied so far in VANET-related literature. In this paper a malicious data detection method is introduced for VANET-based traffic signal control systems. The proposed method integrates a model of expected driver behaviour with position verification in order to detect the malicious data that can degrade the performance of road traffic control at signalized intersections.

\section{Proposed method}

This section introduces an approach, which was intended to detect malicious data in VANET applications for road traffic control at signalized intersections. The considered VANET is composed of vehicle nodes and control nodes that manage traffic signals at intersections. Vehicles are equipped with sensors that collect speed and position data. The collected information is periodically transmitted from vehicles to control nodes. The control nodes use this information for optimizing traffic signals to decrease delay of vehicles and increase capacity of a road network.

In order to detect and filter out malicious data, the control node assigns a trust level to each reported vehicle. The trust levels are updated (decreased or increased) after each data delivery by using the rules discussed later in this section. When making decisions related to changes of traffic signals, the control node takes into account only those data that were collected by vehicles with positive trust level. The data delivered by vehicles with trust level below or equal to zero are recognized as malicious and ignored.

At each time step vehicle reports ID of its current traffic lane, its position along the lane, and velocity. If vehicles $i$ and $j$ are moving in the same lane during some time period and at the beginning of this period vehicle $i$ is in front of vehicle $j$ then the same order of vehicles has to be observed for the entire period. Thus, the following rule is used to detect the unrealistic behaviour of vehicles in a single lane:
\begin{equation}
x_i(t) - x_j(t) > \epsilon_\mathrm{x} \wedge x_i(t-\delta) - x_j(t-\delta) < -\epsilon_\mathrm{x} \wedge l_i(t') = l_j(t') \; \forall t':t-\delta \le t' \le t,
\end{equation}
where: $l_i(t)$, $x_i(t)$, $v_i(t)$ denote respectively lane ID, position, and velocity of vehicle $i$ at time step $t$, $\delta$ is length of the time period, and $\epsilon_\mathrm{x}$ is maximum localization error. It is assumed that the frequency of data reports enables recognition of overtaking. In situation when condition (1) is satisfied and both vehicles have positive trust level, the trust level of both vehicles ($i$ and $j$) is decreased by value $\alpha$ because the collected data do not allow us to recognize which one of the two reported vehicles is malicious. If one of the two vehicles has non-positive trust level then the trust level is decreased only for this vehicle.

According to the second rule (so-called reaction to signal rule), current traffic signals are taken into account in order to recognize the malicious data. If the information received from vehicle $i$ indicates that this vehicle enters an intersection when red signal is displayed or stops at green signal then the trust level of vehicle $i$ is decreased by value $\alpha$. The following condition is used to detect the vehicles passing at red signal:
\begin{equation}
h_n - x_j(t - \delta) > \epsilon_\mathrm{x} \wedge h_n - x_j(t) < -\epsilon_\mathrm{x} \wedge s_n(t') = \mathrm{red} \; \forall t': t - \delta \le t' \le t,
\end{equation}
where: $h_n$ is position of the stop line for signal $n$, $s_n(t)$ is the colour of signal $n$ at time $t$, and the remaining symbols are identical to those defined for rule (1). The vehicles stopped at green signal are recognized according to condition: 	
\begin{equation}
|h_n - x_j(t')| < \epsilon_\mathrm{x} \; \forall t': t - \delta \le t' \le t \wedge s_n(t') = \mathrm{green} \; \forall t': t - \delta \le t' \le t,
\end{equation}
where $|\cdot|$ denotes absolute value and the remaining symbols were defined above. In opposite situations, when the vehicle enters the intersection during green signal or stops at red signal then its trust level is increased by $\alpha$.

Theoretical models of vehicular traffic assume that vehicles move with desired free flow velocity if they are not affected by other vehicles or traffic signals \cite{bbp_22}. Based on this assumption, expected velocity of vehicle $i$ can be estimated as follows: 
\begin{equation}
\hat{v}_i(t) = \min \left( v_f, \frac{h_i(t) - h_{\min}}{\tau} \right),
\end{equation}
where: $v_f$ is free flow velocity, $h_i(t)$ is headway distance, i.e., distance between vehicle $i$ and vehicle in front in the same lane or distance between vehicle $i$ and the nearest red signal, $h_\mathrm{min}$ is minimum required headway distance for stopped vehicle, $\tau$ denotes time which is necessary to stop vehicle safely and leave adequate space from the preceding car or traffic signal. Time $\tau$ can be determined according to the two seconds rule, which is suggested by road safety authorities \cite{bbp_23}.

When the velocity reported by a vehicle differs significantly from the expected velocity then the vehicle is suspected to be malicious. Thus, the trust level of the vehicle is decreased. In opposite situation, when the reported velocity is close to the expected value, the trust level is increased. According to the above assumptions, the trust level is updated by adding value $u_i(t) \cdot \beta$ and $u_i(t)$ is calculated using the following formula: 
\begin{equation}
u_i(t)=
\begin{cases}
1, & |\hat{v}_i(t) - v_i(t)| < \epsilon_\mathrm{v}, \\ 
-\frac{|\hat{v}_i(t) - v_i(t)|}{v_f}, & \mathrm{else},
\end{cases}
\end{equation}
where $\epsilon_\mathrm{v}$ is a threshold of the velocity difference, and the remaining symbols were defined earlier. Threshold $\epsilon_\mathrm{v}$ was introduced in Eq. (5) to take into account error of velocity measurement and uncertainty of the expected velocity determination. 

The last rule for updating trust level assumes that vehicles are equipped with sensors which enable detection and localization of neighbouring vehicles within distance $r$. In this case each vehicle reports the information about its own position and speed as well as positions of other vehicles in the neighbourhood. The information about neighbouring vehicles, delivered by vehicle $j$ at time $t$, is represented by set $D_j(t)$:
\begin{equation}
D_j(t) = \left\{ \left< x_k(t), l_k(t)\right> \right\}, 
\end{equation}
where $x_k(t)$ and $l_k(t)$ denote position and lane of $k$-th vehicle in the neighbourhood.
The additional data can be utilized by control node for verification of the collected vehicle positions. Position of vehicle $i$ should correspond to one of the positions of neighbours ($k$) reported by vehicle $j$ if distance between vehicles $i$ and $j$ is not greater than $r$. Therefore, trust level of vehicle $i$ is increased by $\alpha$ if
\begin{equation}
dist(i, j) \le r \wedge \exists \left< x_k(t), l_k(t)\right> \in D_j(t): dist(i, k) \le \epsilon_\mathrm{x}, 
\end{equation}
where: $dist(i, j)$ is distance between vehicles $i$ and $j$, $r$ is localization range, and the remaining symbols were defined earlier. The trust level is decreased by $\alpha$ if
\begin{equation}
dist(i, j) \le r \wedge dist(i, k) > \epsilon_\mathrm{x} \forall \left< x_k(t), l_k(t)\right> \in D_j(t). 
\end{equation}
The symbols in (8) were defined earlier in this section. The above rule is applied only if the trust level of vehicle $j$ is positive.

It should be noted that the parameter of decreasing trust level for the velocity-related rule ($\beta$) is different than for the remaining rules ($\alpha$) because the velocity-related rule can be used after each data transfer, i.e., significantly more frequently than the other rules.

\section{Experiments}

Simulation experiments were performed to evaluate effectiveness of the proposed method for malicious data detection. The experimental results presented in this section concern percentages of detected malicious data and the impact of these data on vehicle delay at signalized intersections in a road network. 

\begin{figure}
\centering
\includegraphics [width=5cm] {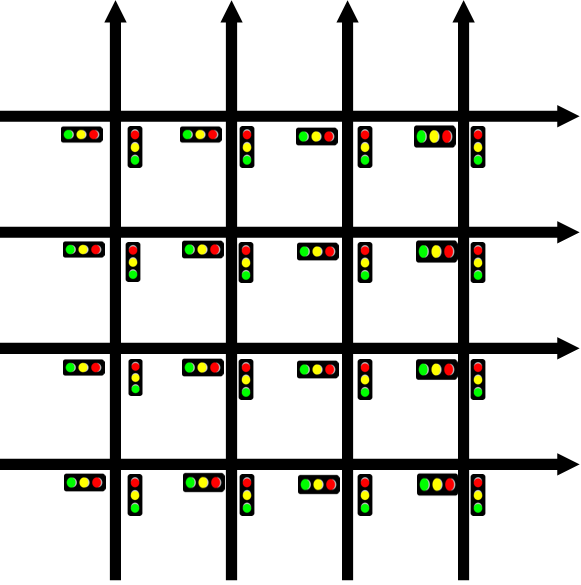}
\caption{Simulated road network}
\label{fig:f1bp}
\end{figure}

In this study the stochastic cellular automata model of road network, proposed by Brockfeld et al. \cite{bbp_24}, was used for the traffic simulation. This model represents a road network in a Manhattan-like city. Topology of the simulated network is a square lattice of 8 unidirectional roads with 16 signalized intersections (Fig.~\ref{fig:f1bp}). The distance between intersections is of 300 m. Each link in the network is represented by a one-dimensional cellular automaton. An occupied cell on the cellular automaton symbolizes a single vehicle. At each discrete time step (1 second) the state of cellular automata is updated according to four steps (acceleration, braking due to other vehicles or traffic light, velocity randomization, and movement). These steps are necessary to reproduce the basic features of real traffic flow. Step 1 represents driver tendency to drive as fast as possible, step 2 is necessary to avoid collisions and step 3 introduces random perturbations necessary to take into account changes of vehicle velocity in regions with high density. Finally, in step 4 the vehicles are moved according to the new velocity calculated in steps 1-3. Steps 1-4 are applied in parallel for all vehicles. Detailed definitions of these steps can be found in \cite{bbp_24}. Maximum velocity was set to 2 cells per second (54 km/h). 

Deceleration probability $p$ for the Brockfeld model is 0.15. The saturation flow at intersections is 1700 vehicles per hour of green time. This model was applied to calculate the stop delay of vehicles. The traffic signals were controlled by using the self-organizing method based on priorities that correspond to "pressures" induced by vehicles waiting at an intersection \cite{bbp_25}. The traffic signal control was simulated assuming the intergreen times of 5 s and the maximum period of 120 s. At each intersection there are two alternative control actions: the green signal can be given to vehicles coming from south or to those that are coming from west.  The simulator was implemented in Matlab.

Intensity of the traffic flow is determined for the network model by parameter $q$ in vehicles per second. This parameter refers to all traffic streams entering the road network. At each time step vehicles are randomly generated with a probability equal to the intensity $q$ in all traffic lanes of the network model. Similarly, the false (malicious) vehicles are generated with intensity $q_\mathrm{F}$ at random locations. The false vehicles move with constant, randomly selected velocity.

During experiments nine algorithms of malicious data detection were taken into account (Tab. 1). The algorithms use different combinations of the rules proposed in Sect. 3 (4 rules used separately and 5 selected combinations that achieved the most promising results in preliminary tests). Simulations were performed for four various intensities of true vehicles $(q = 0.02, 0.06, 0.10, 0.14)$ and false vehicles $(q_\mathrm{F} = 0.02, 0.04, 0.06, 0.08)$. For each combination of the intensities $q$ and $q_\mathrm{F}$ the simulation was executed in 20 runs of 10 minutes. Based on preliminary results, the parameters used for updating the trust levels were set as follows: $\alpha = 1$ and $\beta = 0.2$.

\begin{table}
\centering
\caption{Compared algorithms for malicious data detection}

\begin{tabular}{lccccccccc}
\hline
Algorithm~~~~~~~~~~~~Rule &~~1~~&~~2~~&~~3~~&~~4~~&~~5~~&~~6~~&~~7~~&~~8~~&~~9~~\\
\hline
1 - vehicles order & + & - & - & - & + & - & + & - & + \\
2 - reaction to signals & - & - & + & - & + & + & - & - & + \\
3 - expected velocity & - & - & - & + & - & - & + & + & + \\
4 - neighbour detection & - & + & - & - & - & + & - & + & + \\
\hline 
\end{tabular} 
\label{tab:tab1bp}
\end{table}

\begin{figure}
\centering
\includegraphics [width=12cm] {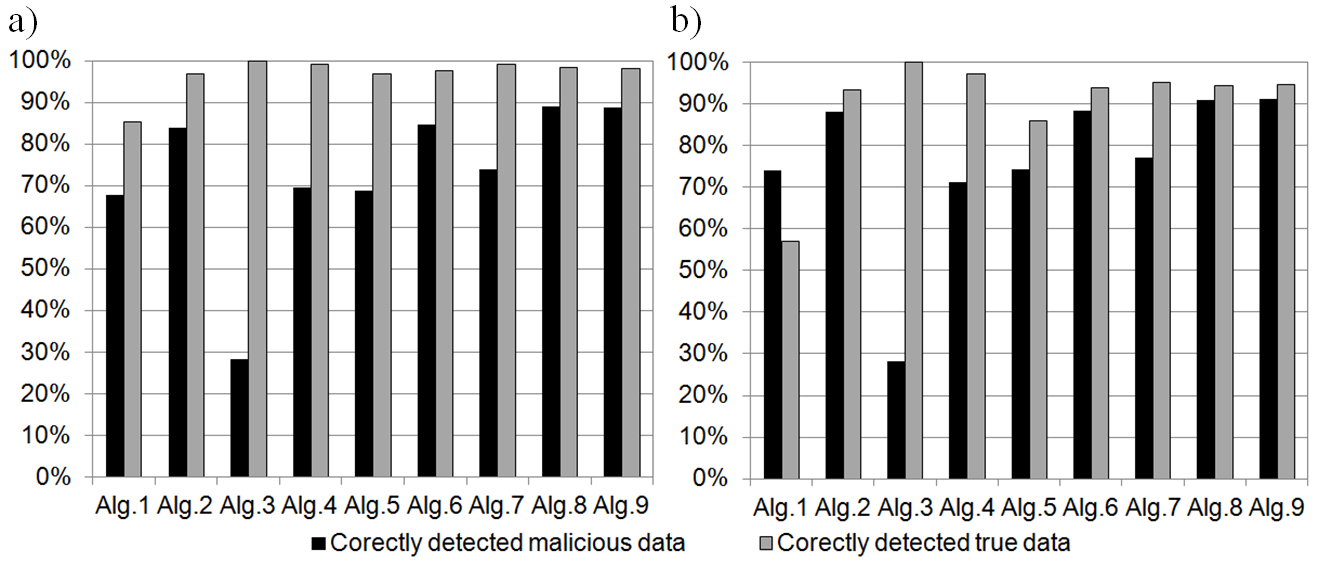}
\caption{Accuracy of malicious data detection for the compared algorithms: a) $q_\mathrm{F} = 0.02$ veh./s, b) $q_\mathrm{F} = 0.08$ veh./s}
\label{fig:f2bp}
\end{figure}

Figure~\ref{fig:f2bp} shows percentages of correctly detected malicious data and correctly recognized true data for two different intensities of false vehicles. The data were categorized as true or malicious at each one-second interval.

\begin{figure}
\centering
\includegraphics [width=8cm] {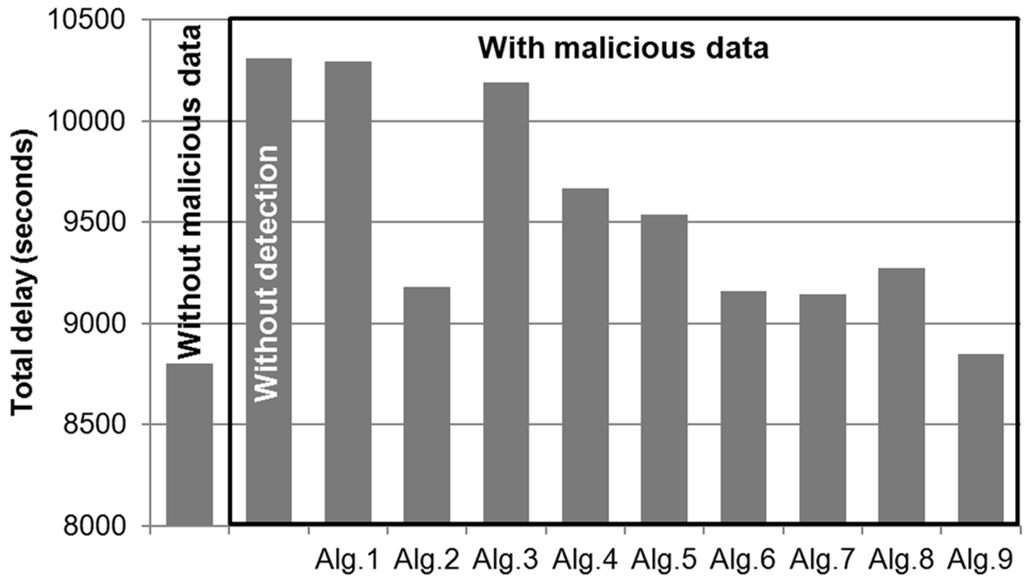}
\caption{Average delay of vehicles for the compared algorithms}
\label{fig:f3bp}
\end{figure}

Total delay of vehicles for the considered algorithms is compared in Fig.~\ref{fig:f3bp}. The results in Fig.~\ref{fig:f3bp} were averaged for all considered true and false vehicle intensities. Average number of vehicles for one simulation run (10 minutes) was 384. The best results were obtained for algorithm 9, which utilizes all proposed rules for detection of the malicious data. This algorithm allows the delay of vehicles to be kept at the low level (close to the value obtained for simulation without malicious data). The delay is increased only by 1\% in comparison to the delay observed when no malicious data are present. Algorithm 9 correctly recognizes 90\% of the malicious data and 96\% of the true data on average. High accuracy was also observed for Algorithm 2, which uses the approach of position verification by neighbouring vehicles without any additional rules. The least satisfactory results were obtained when using the vehicles order rule (Algorithm 1) or the reaction to signals rule (Algorithm 3). For these algorithms the delay of vehicles is close to that observed when the detection of malicious data is not used.

\begin{figure}
\centering
\includegraphics [width=12.2cm] {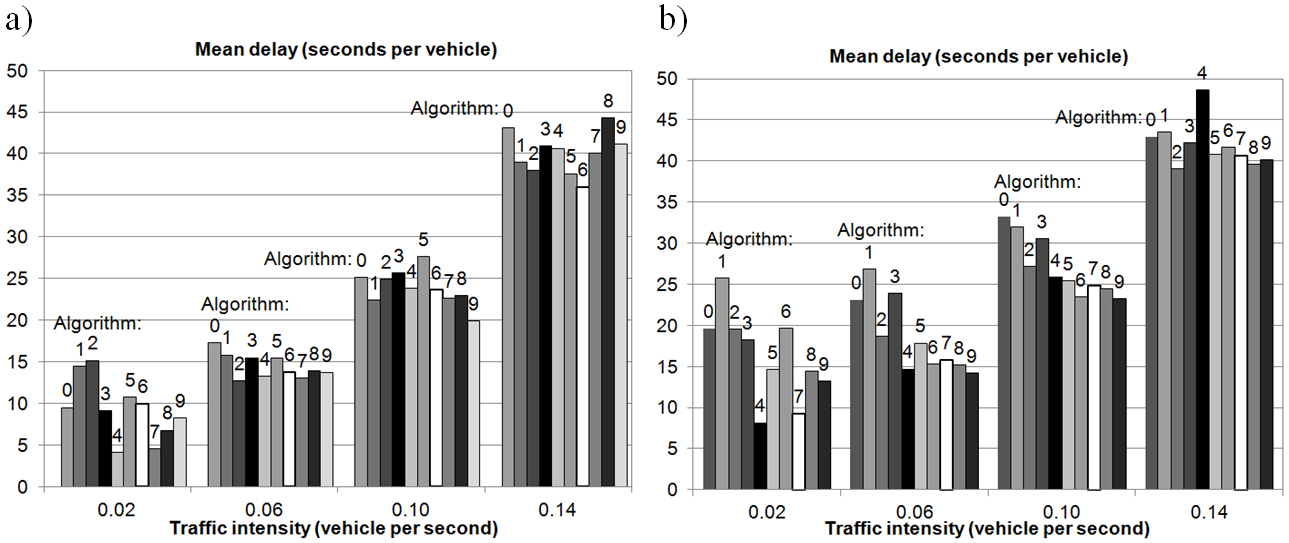}
\caption{Mean delay for different traffic intensities: a) $q_\mathrm{F} = 0.02$ veh./s, b) $q_\mathrm{F} = 0.08$ veh./s}
\label{fig:f4bp}
\end{figure}

Figure~\ref{fig:f4bp} shows mean vehicle delays for various intensities of the traffic flow ($q$) and two different intensities of the false vehicles generation ($q_\mathrm{F}$). Algorithm 0 in Fig.~\ref{fig:f4bp} corresponds to the situation when no malicious data detection is implemented. It can be observed in these results that the effectiveness of a particular algorithm strongly depends on the considered intensities. For instance, in case of $q = 0.14$ and $q_\mathrm{F} = 0.02$  Algorithm 8 causes a higher delay than those obtained without malicious data detection, while for the remaining intensities Algorithm 8 gives good results. However, for Algorithm 9 the delay is reduced when comparing with those obtained without malicious data detection for all considered intensity settings. This fact confirms that all the proposed rules are useful as they contribute with different degree in various traffic conditions to mitigating the negative impact of malicious data.

\section{Conclusion}

Sybil attacks can degrade the performance of VANET-based traffic signal control. The proposed approach enables effective detection of malicious data created in VANETs when the Sybil attacks are launched. The introduced detection scheme is based on rules that take into account unrealistic overtaking manoeuvres, expected driver behaviour (reaction to traffic signals and preferred velocity) as well as verification of vehicle position by neighbouring nodes. Effectiveness of this approach was demonstrated in simulation experiments for a decentralized self-organizing system that controls the traffic signals at multiple intersections in an urban road network. The experimental results show that combination of different detection mechanisms allows the malicious data in VANET to be correctly recognized and is essential for mitigating their negative impact on the performance of traffic signal control. Further research is necessary to integrate the method with more sophisticated models of driver behaviours, enable automatic parameters calibration based on collected data, and test the proposed approach in different (more realistic) scenarios with various traffic control algorithms.

\end{document}